\documentstyle[12pt]{article}
\begin{document}
\vspace{1.0in}
\begin{flushright}
{\small February 24, 1993}
\end{flushright}
\vspace{2.0cm}
\begin{center}
{\Large{\bf Modeling Complex Nuclear Spectra\\
- Regularity versus Chaos}}
\vspace{1.0cm}

S. Dro\.zd\.z$^{a,b}$,
S. Nishizaki$^{a,c}$,
J. Speth$^{a}$\footnote{also at: Institut f\"ur Theoretische Kernphysik,
Universit\"at Bonn, D-5300 Bonn, Germany} and
J. Wambach$^{a,d}$
\vspace{1.0cm}

{\it
a) Institut f\"ur Kernphysik, Forschungszentrum J\"ulich,
D-5170 J\"ulich, Germany \newline
 b) Institute of Nuclear Physics, PL - 31-342 Krak\'ow, Poland \newline
 c) College of Humanities
and Social Sciences, Iwate University, Ueda  3-18-34,\newline
Morioka 020, Japan\newline
 d) Department of Physics, University of Illinois at Urbana,
IL 61801, USA}
\end{center}
\vskip 1cm
\abstract
{A statistical analysis of the spectrum of two particle - two hole
doorway states in a finite nucleus is performed.
On the unperturbed mean-field level sizable attractive correlations
are present in such a spectrum. Including  particle-hole
rescattering effects via the residual interaction
introduces repulsive dynamical correlations which generate the
fluctuation properties characteristic of the Gaussian Orthogonal
Ensemble. This signals that the underlying dynamics becomes chaotic.
This feature turns out to be independent of the detailed form
of the residual interaction and hence  reflects the generic nature
of the fluctuations studied.}
\vspace{1.0cm}

\newpage

Theoretical models aiming at the quantal description of an excitation
and subsequent decay of a collective state in a many-body system
are usually based on a
division of the full Hilbert space into two sectors, $S_1$ and $S_2$,
spanned by the vectors $|1\rangle$ and $|2\rangle$, respectively.
At the same time, the hamiltonian is represented as
${\hat H}={\hat H_o} + {\hat V}$, such that
$\langle 1|{\hat H}_o|1'\rangle = \epsilon^o_1 \delta_{11'}$,
$\langle 2|{\hat H}_o|2'\rangle = \epsilon^o_2 \delta_{22'}$ and
$\langle 1|{\hat H}_o|2 \rangle = 0$. When $\hat V$ is taken
into account the above relations no longer hold.
By diagonalizing $\hat H$ in the basis $|2 \rangle$ and redefining
$|2 \rangle$ one can still have
$\langle 2|{\hat H}|2'\rangle = \epsilon_2 \delta_{22'}$, however.

A collective state $|f \rangle$ such as a plasmon or a nuclear
vibrational mode is defined as an eigenstate of
$\hat H$ in the subspace $S_1$:
$|f \rangle = \sum_1 f_1 |1 \rangle$.
In the full space, including $S_1$ and $S_2$, $|f \rangle$
is, however, no longer an eigenstate of $\hat H$ but rather a wave
packet which begins to 'leak' into the space $S_2$. This
constitutes a mechanism for dissipation.
Thus, for the time-dependent state one has
\begin{equation}
|f(t) \rangle = \sum_1 f_1(t) |1 \rangle +
\sum_2 f_2(t)|2 \rangle,\qquad (f_2(t=0)=0)
\end{equation}
and the Schr\"odinger equation for $f_1$ and $f_2$
reads:
\begin{equation}
i{d\over dt} \left\lbrack\matrix{f_1\cr f_2}\right\rbrack =
\left\lbrack\matrix{ H_{11'}&H_{12'}\cr H_{21'}&H_{22'}\cr}
\right\rbrack
\left\lbrack\matrix{f_{1'}\cr f_{2'}}\right\rbrack.
\end{equation}
Applying a procedure similar to the Nakajima-Zwanzig projection
technique \cite{NZ}, $i.e.$, solving the second of those
equations  for $f_2(t)$, inserting into the first one for $f_1(t)$
and assuming that the basis $|2\rangle$ is already defined
such that $\langle 2|\hat H| 2' \rangle = \epsilon_2 \delta_{22'}$,
yields
\begin{equation}
i{d\over dt} f_1(t) - \sum_{1'} H_{11'} f_{1'} (t) =
\sum_{1'} \int_0^t d\tau f_{1'}(t-\tau) v_{11'} (\tau),
\label{eq:f1}
\end{equation}
where
\begin{equation}
v_{11'}(\tau) = -i\sum_2 H_{12} H_{21'}
\exp (-i\epsilon_2 \tau).
\label{eq:v1}
\end{equation}
The $r.h.s.$ in eq.~(\ref{eq:f1}) is a generalized non-local
collision term
which compensates for the formal elimination of the space $S_2$.
The matrix elements $H_{12} = \langle 1|\hat H|2 \rangle$ describe
the degree of mixing between $S_1$ and $S_2$ after
inclusion of $\hat V$.
The internal structure of space $S_2$ is represented by the eigenvalues
$\epsilon_2$ of the full hamiltonian $\hat H$ in this space.

In nuclear physics the space $S_1$ is typically spanned by
one particle - one hole (1p1h) states generated from
the mean field hamiltonian, $\hat H_o$. Diagonalizing
the full nuclear hamiltonian in such a basis gives the
RPA boson excitations. Ideally, the space $S_2$, absorbing such
an excitation, should contain all possible $n$p$n$h states.
Then, after diagonalization of $\hat H$, the space $S_2$
would reflect the entire  complexity of the spectrum of a compound
nucleus. This, however, is neither possible for practical reasons
(too many states) nor does it seem necessary for physical reasons.
The nuclear hamiltonian involves predominantly two-body interactions
and thus couples the 1p1h RPA bosons chiefly to the 2p2h
components in the space $S_2$. In this way, the 2p2h states play
a special role of 'doorway states' \cite{Wam}.
For these reasons, the space $S_2$
is usually specified in terms of the 2p2h states alone \cite{DNS}.
The resulting methods are known as extended RPA
and are believed to account for the spreading width of collective
modes.
Moreover, since the spectrum of such states is already very dense
in energy, the interaction in the 2p2h - space is usually
neglected which
corresponds to using the unperturbed 2p2h energies  $\epsilon^o_2$
as $\epsilon_2$ in eq.~(\ref{eq:v1}).

At this point one should recall that 'nuclear dissipation is a process
occurring in a closed system, and very likely a consequence of chaotic
motion' \cite{Wei1}. Evidence supporting the connection between
dissipation and chaos comes both, from classical \cite{SSD}
and from quantum \cite{BRW} considerations.
In light of the introductory remarks such a link seems indeed to be
present.
The fluctuation properties of the spectra of compound nuclei are
consistent \cite{HPB} with those of the Gaussian Orthogonal
Ensemble (GOE) of random matrices \cite{BFF}.
The same fluctuation properties are identified theoretically
\cite{BGS} as well as experimentally \cite{SS}
for those quantum systems  whose classical counterparts are chaotic.
Similar conclusions can be drawn from the study of open phase
space, scattering  phenomena \cite{BS}.
It is thus natural to require that modeling the space $S_2$
should preserve this fundamental property of the compound nucleus.
Because of the special role played by the 2p2h states,
the purpose of the present letter is to quantitatively explore the
problem of whether and under which conditions already
the basis of such states can support the GOE fluctuation
characteristics and thus makes the underlying dynamics chaotic.

To make our analysis as realistic as possible
the mean field hamiltonian $\hat H_o$,
generating the single-particle states,
is specified in terms of a local Woods-Saxon potential \cite{BM}:
\begin{equation}
U(r) = - {V_o\over{1+e^{{r-r_a}\over a}}} - V_{LS}
{1\over r}{d\over dr}({1\over {1+e^{{r-r_a}\over a}}})
(\vec l\cdot\vec\sigma)+V_c(r)
\label{eq:WS}
\end{equation}
where $V_c(r)$ is the Coulomb potential. The parameters $V_o$,
$V_{LS}$ and $a$ are adjusted such as to reproduce the ground state
properties of the nucleus and, in particular, the single-particle
energies near the Fermi surface. To determine the single-particle
wave functions from this potential, $\hat H_o$ is diagonalized
in a harmonic oscillator basis including 16 major shells.

As a residual interaction $V$ we adopt the zero-range
Landau-Migdal interaction
\begin{equation}
V(r_1,r_2)= C_o ~(f+f^\prime\tau_1\cdot\tau_2+
g\sigma_1\cdot\sigma_2+g^\prime\sigma_1\cdot\sigma_2\tau_1\cdot\tau_2)
\delta(r_1-r_2)
\label{eq:LM}
\end{equation}
which provides a good description of low-energy nuclear excitations.
In extended RPA calculations with this interaction
(including unperturbed 2p2h states only)
the basis of single-particle states is
typically specified \cite{SW} by four major shells, two below
and two above the Fermi surface.
Consistently, in the present study we use the basis
of similar size.
We employ the set of empirical Landau-Migdal parameters as given
in ref.\cite{Tow} $(f=-0.1, f'=0.6, g=0.15$ and $g'=0.7$
with $C_o=392$MeV fm$^3$)
and perform the calculations for $^{40}$Ca.

In spherical nuclei, such as $^{40}$Ca, the total angular momentum $J$
and parity $\pi$ are good quantum numbers which determines the selection
of the basic 2p2h states, active in the decay of a given RPA boson
excitation. All three types of the matrix elements generated by the
residual two-body interaction in the basis of 2p2h states are shown
diagrammatically in Fig.~1 and the angular momentum
coupling scheme is also indicated.
The single-particle basis of four major shells in the Woods-Saxon
potential allows
realistic estimates \cite{SW} of global spreading widths because
the number of the corresponding 2p2h states is already of the order
of $10^3$ - $10^4$, depending on the multipolarity of an excitation
with the maximum of 11720 for $J^{\pi}=3^-$.
The matrix of such a size cannot, however, be diagonalized
with satisfactory precision. A manageable
number (2696) of 2p2h states is found for $J^{\pi}=0^+$ and,
therefore, our further discussion will be limited to this multipole.
The states are not coupled to the good isospin $T$ because the
hamiltonian, as specified above, explicitly violates the isospin
symmetry. An even weak violation of this symmetry is known \cite{MBE}
to correlate the $T=0$ and $T=1$ states which allows to combine the
relevant spectra for a statistical analysis.

The most obvious measure of spectral fluctuations is the one
expressed in terms of the nearest-neighbor spacing (NNS)
distribution. A standard procedure of analysis is to normalize
the spectrum such that the fluctuations on different energy scales
are directly comparable. Here we perform the corresponding unfolding
\cite{BG} by approximating the integrated density of states
with polynomials up to the order of 12. This guarantees stability
of the result.
Before unfolding we discard the 50 lowest states in order
to eliminate those states which are located far from their
main concentration.
Similarly we discard the 50 highest-lying states.

We find three qualitatively different situations as
illustrated in Fig. 2. Part (a) shows the NNS distribution for the
spectrum of unperturbed 2p2h states coupled to $J^{\pi}=0^+$.
Such a spectrum is not generic and is characteristic for a narrow class
of integrable systems involving extra correlations
in the hamiltonian \cite{BT}.
The pronounced peak for small nearest-neighbor separations
illustrates a strong tendency of states for clustering.
Actually, in most cases these are even exact degeneracies.
They reflect the fact that within the single-particle basis used
there are many more two hole (hh) states than the number of different
energies available. Analogous partitioning
is even more restrictive on the two particle (pp) side.
Including the interaction in the pp (diagram (a) in Fig.~1)
and hh (diagram (c))
channels removes those degeneracies and, as shown in part (b)
of Fig.~2, immediately brings the spectrum to the known universality
class of generic integrable systems \cite{BT} characterized by a
completely uncorrelated sequence of eigenenergies. As a consequence,
the NNS follows a Poisson distribution. This result essentially does
not depend on precise values of the parameters of
the residual interaction.
Already, a comparatively weak perturbation
(for instance 20$\%$ of the original strength)
produces a similar picture.
The other important element which makes the spectrum
uncorrelated is that, at this level, the system considered (2p2h)
still remains a simple product of two subsystems (pp and hh)
and the corresponding energies are the sums of two independent
components. The crucial step which brings us
to the situation displayed in part (c)
of Fig.~2 is the inclusion of the ph-type matrix elements, represented
by the diagram (b) of Fig.~1.
These matrix elements introduce such important correlations that
the resulting NNS distribution almost perfectly follows a Wigner
distribution, i.e. is consistent with the GOE. In fact, including only
this diagram and ignoring the other two ((a) and (c)) gives
the same result. Again this feature does not depend significantly
on the strength of the interaction parameters. Varying them in
a broad range (factors between 1/5 and 2 for instance) leaves
the histogram in Fig.~2.c essentially unchanged. This is another
confirmation that the fluctuations, which we are looking at,
reflect the generic properties of the system.

The discussion presented above is based on the subspace of 2p2h states
coupled to $J^{\pi}=0^+$. Similar behavior can, however, be expected
for the other multipoles.
In smaller model spaces where the 2p2h diagonalization can be done
reliably for any multipole we see no qualitative
difference when performing a similar study.
Of course, in the smaller spaces, the mean square deviations are larger
due to poorer statistics.

Because of the similarity of wave functions, the most important
doorway states for decay of a typical giant resonance
are those 2p2h states which are generated by the
single-particle basis of two major mean-field shells on
both sides of the Fermi surface.
Those states have, therefore, been discussed extensively
above. To conclude about the degree of genuine chaoticity of the
nuclear hamiltonian projected onto the full 2p2h space one needs,
however, a more restrictive test.
The point is that including further single-particle shells
in the diagonalization may influence the higher-energy part of the
spectrum studied above. Since such an extension cannot be made in
practice we go the opposite way. Out of the total of 2696
$J^{\pi}=0^+$ states we select a sequence of 400 states starting
from 51-st state.
Such a sequence covers the excitation energies up to about 30 MeV.
Concerning the overlap with the high-energy 2p2h states, not
taken into account in the diagonalization, we find this choice safe.
The resulting NNS distribution and the $\Delta_3$ statistics
\cite{DM} the latter being a measure of the rigidity of the spectrum
are shown in Fig.~3, including all diagrams of Fig.~1.
The $\Delta_3(L)$ statistics is calculated as an average of the
mean-square deviation from the straight line of the integrated density
of (unfolded) states in the intervals of length $L$.
A comparison with
the corresponding Poisson and GOE predictions \cite{BFF} is also
made. We have verified that
using an analogous sequence of states on the unperturbed level
reveals a similar clustering as in Fig.~2.a. Furthermore,
including  diagrams (a) or (c) or both leads to decorrelation
seen already for a string of eigenvalues of the same length.
The presence of such a transition means that the final
good agreement with the GOE does not involve any kinematical
repulsion \cite{DS}, but is entirely due to the dynamical
correlations violating integrability and thus generating chaos.

In conclusion, already on the 2p2h level the spectral fluctuations
of the GOE can manifest themselves, provided the residual interaction
is taken into account. Most important in this connection are
the particle-hole rescattering effects which correlate the states
from both sides of the Fermi surface. While the unperturbed 2p2h
states certainly provide a reasonable first approximation for the
global energy location of the doorway states, they fail completely
in the sense of fluctuations. This may have an influence on the
fine structure of a resonance.
The above aspect of nuclear dynamics should also be kept in mind
when addressing the question \cite{HSY} as to whether the nuclear
collective motion is Markovian or not \cite{ADY}.
In classical terms a chaotic system looses memory very fast because of
exponential instabilities.  It is thus natural to expect that chaos
suppresses the role of the history also on the quantum level.
Actually, the integration kernel in eq.~(\ref{eq:f1}) involves a sum of
exponents eq.~(\ref{eq:v1}) over the whole background spectrum.
For a spectrum with a strong tendency to clustering many terms
in such a sum may add up constructively and thus amplify the $\tau$
dependence in eq.~(\ref{eq:v1}). The opposite should apply to
the GOE spectrum.
In fact, a simple estimate of such effects based on the spectra discussed
in this letter shows one order of magnitude reduction of the amplitude
of oscillations when going from the case of Fig.~2.a to Fig.~2.c.
This is an important problem \cite{Pri}
which demands a more systematic study.

We thank J.J.M. Verbaarschot for very helpful conversations.
This work was supported in part by the Polish Committee of
Scientific Research
and by NSF grant PHY-89-21025. One of the authors (S. N.) would like to
express his thanks to the Alexander von Humboldt Foundation for a
fellowship.

\newpage
\vspace{.25in}
\parindent=.0cm            

\newpage
\begin{center}
{\bf Figure Captions}
\end{center}

\noindent
Fig.~1 :Diagrammatic representation of the two-body matrix
elements in the space of 2p2h excitations with explicit indication
of the angular momentum coupling scheme.

\vspace{0.20in}
\noindent
Fig.~2 : Nearest-neighbor spacing distributions (histograms)
for the sequence of 2p2h states coupled to $J^{\pi}=0^+$ as
a function of the normalized relative distance $s$. The distribution
has been generated in a  basis of four major shells
(two below the Fermi surface and two above)
for a Woods-Saxon potential with  parameters corresponding to
$^{40}$Ca.
Part (a) displays the unperturbed case, (b) corresponds to  the results
from a diagonalization including diagrams (a) and (c) of Fig. 1
and (c) includes all the diagrams.
The residual interaction used is given by eq.~(\ref{eq:LM}).
The dotted lines represent the Poisson and the solid lines
the Wigner distributions.

\vspace{0.20in}
\noindent
Fig.~3 : Nearest-neighbor spacing distribution
(histogram in the upper part)
and the $\Delta_3$ statistics (diamonds in the
lower part) for a sequence of
400 low-energy states (between 51 and 450)
 obtained from a  diagonalization of the residual interaction
eq.~(\ref{eq:LM}) in the basis of 2p2h states for $J^{\pi}=0^+$.
The dotted lines refer to the Poissonian spectrum and the
solid lines to GOE predictions.

\end{document}